\documentclass[debug]{rmaa}

\usepackage{paralist}

\usepackage{psfrag,color}

\usepackage[latin1]{inputenc}

\title{Asymmetric Shapes of Radio Recombination Lines from 
Ionized Stellar Winds} 

\author{
  R. Ignace \altaffilmark{1} }

\altaffiltext{1}{Department of Physics \& Astronomy, East Tennessee
  State University.}

\shortauthor{Ignace}
\shorttitle{Asymmetric Radio Recombination Lines}

\fulladdresses{
\item Richard Ignace: Department of Physics \& Astronomy, East Tennessee
  State University, Johnson City, TN 37614, USA (ignace@etsu.edu).}

\listofauthors{Richard Ignace}
\indexauthor{Ignace, R.}

\abstract{
Recombination line profile shapes are derived for ionized spherical
stellar winds at radio wavelengths.  It is assumed that the wind
is optically thick owing to free-free opacity.  Emission lines of
arbitrary optical depth are obtained assuming that the free-free
photosphere forms in the outer, constant expansion portion of the
wind.  Previous works have derived analytic results for isothermal winds
when the line and continuum source functions are equal.  Here,
semi-analytic results are derived for when the source functions are
not equal to reveal that line shapes can be asymmetric about line
center.  A parameter study is presented and applications discussed.
  }

\resumen{
  }

\addkeyword{line: profiles}
\addkeyword{radiative transfer}
\addkeyword{stars: early-type}
\addkeyword{stars: winds, outflows}
\addkeyword{radio lines: stars}

\begin{document}
\maketitle

\section{Introduction}
\label{sec:intro}

Radio astronomy has long proven to be an important window into the study
of stellar astrophysics, and stellar outflows have been no exception
(e.g., Dulk 1995; G\"{u}del 2002; Kurt et al.\ 2002).  For stellar winds
a key driver has been the prospect of measuring wind mass-loss rates,
$\dot{M}$, from the excess infrared (IR) and radio continuum emission
relative to the stellar atmosphere (e.g., Panagia \& Felli 1975; Wright
\& Barlow 1975).  Numerous studies have focused on determining $\dot{M}$
values based on this approach (e.g., Abbott et al.\ 1980; Abbot, Bieging,
\& Churchwell 1981; Abbott et al.\ 1986; Bieging, Abbott, \& Churchwell
1989; Leitherer, Chapman, \& Koribalski 1995).

One of the main results from a consideration of free-free excesses
formed in the wind is that the spectral energy distribution (SED)
at long wavelengths will have a power-law slope with flux $f_\nu \propto
\lambda^{-0.6}$.  However, this outcome depends on several assumptions:
isothermal, spherical symmetry, large optical depth, negligible
contribution from the stellar atmosphere, and constant outflow speed.
Cassinelli \& Hartmann (1977) explored the effects of different power
laws for the wind density and temperature distributions to relate
the SED power-law slope to these influences.  Schmid-Burgk (1982)
showed that such SED slopes persist even for axisymmetric stellar
envelopes, as long as the same power-law relations are adopted.  The
main difference is that flux levels are modified, which would have
implications for inferring $\dot{M}$ values.

Of greater relevance in recent decades has been the abundance of evidence
for clumping in massive star winds.  In this regard the literature is
voluminous, and there has even been a conference to focus on the topic
(Hamann, Feldmeier, \& Oskinova 2008).  The line-driven winds of
massive stars (Castor, Abbot, \& Klein 1975; Friend \& Abbott 1986;
Pauldrach, Puls, \& Kudritzski 1986) are known to be subject to an
instability (e.g., Lucy \& White 1980; Owocki, Castor, \& Rybicki 1988).
This instability produces shocks in the flow and is a natural culprit
for stochastic wind clumping. The clumping is well-known to affect the
long wavelength emission because of the density-square dependence of the
free-free emissivity.  In the presence of clumping, the radio emission
is overly bright for a given value of $\dot{M}$ as compared to a smooth
(i.e., unclumped) wind with the same mass loss.  Neglecting the clumping
leads to overestimates of $\dot{M}$, scaling as the square root of the
clumping factor, or inverse to the square root of the volume filling
factor of clumps.  These factors will be defined precisely in the
following section.

Clumping affects any density-square emissivity, including recombination
lines.  Clumping has been incorporated into several detailed complex
numerical codes for modeling massive star atmospheres and their winds, such
as CMFGEN (Hillier \& Miller 1999) and PoWR (Hamann, Gr\"{a}fener, \&
Liermann 2006).  An important distinction for clumping
is between ``macroclumping'' and ``microclumping''.  The former
leads to modifications of observables that can depend on the shape
of the clump and is sometimes synonymous with a ``porosity''
treatment.  The latter is when clumps are all optically thin, so
that the radiative transfer does not depend on details of clump
morphology.  Consequently, microclumping can be handled in terms
of a scale parameter, and in fact does not alter the SED slope
relative to a unclumped wind (Nugis, Crowther, \& Willis 1998).
Ignace (2016a) considered the impact of macroclumping vs microclumping
for ionized winds at long wavelength.

This contribution is concerned with modeling a radio recombination line
(RRL) profile shape that also includes continuum free-free opacity.
The problem has been addressed many times before.  Rodr\'{i}guez (1982)
derived the line profile shape for this case, with the interest of
supplementing the use of the continuum to obtain $\dot{M}$ with line
broadening formed in the same spatial locale to obtain the wind terminal
speed $v_\infty$.  Hillier, Jones, \& Hyland (1983) did so as well.
Ignace (2009) repeated the derivation, and expanded the consideration for
inclusion of line blends.  All of these treatments assume that the source
function for the line and continuum is the same, as given by the Planck
function for an isothermal wind.  Using a numerical radiative transfer
calculation, Viner, Vallee, \& Hughes (1979) showed that an asymmetric
line shape can result when the line and continuum source functions
are unequal.  Here, this result is explored further through analytic
derivations.  Section~\ref{sec:modeling} introduces the model assumptions
and presents a derivation for the line shape.  Unlike most previous
treatments, the derivation also allows for a power-law distribution
of microclumping in the wind.  Section~\ref{sec:results} provides for
a parameter study for line profile shapes.  Section~\ref{sec:concs}
discusses relevant applications for various astrophysical sources.

\section{Radio Recombination Line Modeling}
\label{sec:modeling}

Various authors have addressed the relevance of non-LTE effects for
interpreting observed RRLs.  A discussion of progress on the topic can
be found in Gordon \& Sorochenko (2002).  Relevant to wind-broadened
emission lines, Viner et~al.\ undertook a calculation of departure
coefficients for studies of H~{\sc ii regions.  As previously noted,
they allowed for spherical outflow and found that line shapes can be
asymmetric.  Peters, Longmore, \& Dullemond (2012) conducted a similar
study for H~{\sc ii} regions, and elaborated further on line asymmetry for
an outflow.  However, neither Viner et~al.\ nor Peters et~al.\ explored
the possibility of analytic solutions for the radiative transfer.}  Here,
the approach largely follows Ignace (2009), but relaxing the assumption
that the line and continuum source functions are equal.  The primary
assumptions of the model are as follows:

\begin{enumerate}[i.]

\item The wind is spherically symmetric in time average.

\item The wind is optically thick to free-free opacity.  The line
can be thin or thick.

\item While the line and continuum source functions may not be equal,
they are taken as constant with radius.

\item Microclumping is included in the treatment, specifically as a
power-law distribution\footnote{The additional power-law distribution
need not be attributed to clumping.  It could be attributed to something
else that modifies the density.  However, it cannot be the velocity law,
since that would lead to a different geometry for the isovelocity zones
and would invalidate the derivation that follows.  The inclusion of the
additional power law follows the spirit of the approach in Cassinelli
\& Hartmann (1977).} with radius.  Clumping in massive star winds
is both predicted and measured to vary with radius (e.g., Runacres \&
Owocki 2002; Blomme et al.\ 2002, 2003; Puls et al.\ 2006).

\end{enumerate}

\begin{figure}[!t]
\includegraphics[width=\columnwidth]{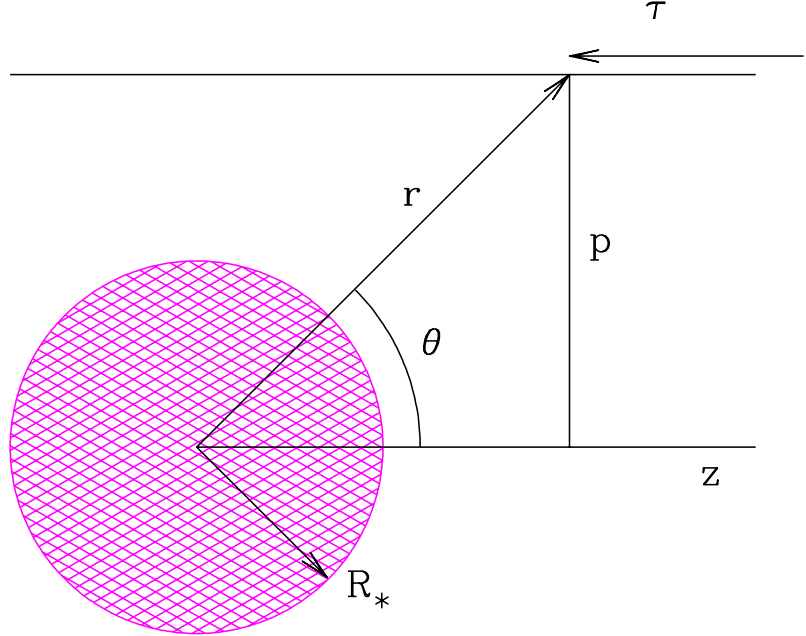}
\caption{Coordinate definitions used in the derivation for
the line and continuum emission from a spherical ionized wind.
See text for explanation.
}
\label{fig1}
\end{figure}

\subsection{Wind Parameters}
\label{sub:wind}

Spherical symmetry requires that the wind has a strictly radial
wind velocity and density.  Being optically thick to free-free opacity, only
the large radius flow at constant expansion will be considered.
The wind terminal speed is represented by $v_\infty$.  The wind
also has a time-average mass-loss rate of $\dot{M}$, and the star
has radius $R_\ast$.

Microclumping is represented as

\begin{equation}
\langle \rho^2 \rangle = D_{\rm cl}\, \langle \rho \rangle^2,
\end{equation}

\noindent where $\langle x \rangle$ represents spatial averaging.
On the right-hand side, the average density is given by the smooth
wind relation for spherical symmetry, with

\begin{equation}
\langle \rho \rangle = \frac{\dot{M}}{4\pi\,R_\ast^2\,v_\infty}\,
	\left(\frac{R_\ast}{r}\right)^2 \equiv \rho_0\,
	\left(\frac{R_\ast}{r}\right)^2.
\end{equation}

\noindent For emissivity $j_\nu\propto \rho^2$, the emission is
enhanced above the smooth wind by the clumping factor $D_{\rm cl}$.
It is often common to represent the clumping in the wind with a volume
filling factor, $f_V=D_{\rm cl}^{-1}$.  Both approaches are used in the
literature (c.f., clumping factor:  Hamann \& Koesterke 1998 or Ignace,
Quigley, \& Cassinelli 2003; volume filling factor:  Abbott et al.\
1981 or Dessart et al.\ 2000).

For this study the clumping factor is allowed to vary with radius as
a power law, with

\begin{equation}
D_{\rm cl} \propto r^{-m}.
\end{equation}

\noindent The case of $m=0$ is for clumping that is constant throughout
the flow; $m>0$ implies that clumping declines with radius; $m<0$ is
the opposite case.  (Note that some care must be taken with use
of the power law for clumping, since $D_{\rm c} \ge 1$.)

\subsection{Line and Continuum Opacities}
\label{sub:opacities}

The free-free opacity, $\kappa_\nu$, is given by

\begin{equation}
\kappa_\nu\,\rho = K_{\rm ff}\,n_{\rm i}\,n_{\rm e},
\end{equation}

\noindent where $n_{\rm i}=\rho/\mu_{\rm i}m_H$ is the number density
of ions, with $\mu_{\rm i}$ the mean molecular weight per free ion;
$n_{\rm e} = \rho/ \mu_{\rm e}m_H$ is the number density of electrons,
with $\mu_{\rm e}$ the mean molecular weight per free electron;
$m_H$ is the mass of hydrogen; and (Cox 2001)

\begin{equation}
K_{\rm ff} = 3.692\times 10^8 \,\left( 1-e^{-h\nu/kT_C}\right)
        Z_{\rm i}^2 g_\nu T_C^{-1/2}\nu^{-3}.
\end{equation}

\noindent In the preceding equation, $h$ is Planck's constant, $k$
is the Boltzmann constant, $T_C$ is the temperature of the gas
appropriate for the continuum emission, $Z_{\rm i}$ is the root
mean square ion charge, $\nu$ is frequency, and $g_\nu$ is the Gaunt
factor.

Figure~\ref{fig1} shows the geometry for evaluating the optical
depth $\tau$ along a ray.  Cylindrical coordinates for the observer
are $(p,\alpha,z)$, with the observer located at great distance along
the $+z$-axis.  Spherical observer coordinates are $(r,\theta,\alpha)$,
with $r^2=p^2+z^2$.  The continuum optical depth along a ray of fixed
impact parameter, $p$, is

\begin{equation}
\tau_C = {\cal T}_C(\lambda)\,\int \tilde{\rho}^2(\tilde{r})\, 
	D_{\rm cl}(\tilde{r})\, d\tilde{z},
\end{equation}

\noindent where $\tilde{x}$ signifies a normalized parameter,
in this case $\tilde{\rho} = \rho/\rho_0$ and lengths are relative
to $R_\ast$, and the optical depth
scaling is

\begin{equation}
{\cal T}_C = \frac{K_{\rm ff}\,R_\ast\,\rho_0^2}
	{\mu_{\rm i}\,\mu_{\rm e}\,m_H^2}.
\end{equation}

\noindent At long wavelengths that are the focus of this paper,
${\cal T}_C \propto g_\nu\,\lambda^2$ for the Rayleigh-Jeans
limit, and $g_\nu \propto \lambda^{0.1}$.

The line opacity is somewhat similar to that of the continuum
in the sense that there is a dependence on the square of density
for recombination.  Assuming that the wind speed is highly 
supersonic, the line optical depth can be approximated from
Sobolev theory (Sobolev 1960).  The line optical depth becomes

\begin{equation}
\tau_L = \frac{\kappa_L\,\rho\,\lambda}{(v_{\infty}/r)\,(1-\mu^2)},
\end{equation}

\noindent where $\kappa_L\,\rho \propto D_{\rm cl}\,\rho^2\,F(T_L)$,
for $F(T_L)$ a function of temperature appropriate for the line 
emission, and $\mu=\cos\theta$.  

For the case of constant expansion of
the wind at $v_{\infty}$, the line-of-sight velocity shift due
to the Doppler effect is $v_{\rm z} = - v_\infty\,\mu$.  It is
convenient to introduce a normalized velocity shift with

\begin{equation}
w_{\rm z} = v_{\rm z}/v_\infty = - \mu.
\end{equation}

\noindent Also note that $p = r\sin\theta$.  Then 
the line optical depth becomes

\begin{equation}
\tau_L = {\cal T}_L\,\tilde{p}^{-3-m}\,(\sin\theta)^{1+m},
\end{equation}

\noindent where the power-law dependence of $D_{\rm cl}$ with radius
has been substituted into the expression, along with $\tilde{\rho}^2
= \tilde{r}^{-4}$, and ${\cal T}_L$ is the optical depth scaling
for the line.  Casting the line optical depth in terms of $p$ and
$\theta$ will prove useful for solving the radiative transfer
problems in the following sections.

\subsection{Solution for the Case of $S_L=S_C$}
\label{sub:same}

When the line and continuum source functions are the same, 
let $S_0 = S_L = S_C$.  At wavelength $\lambda$, just outside
the maximum velocity shift of the line, the flux of continuum 
emission is given by

\begin{equation}
f_C = \frac{2\pi R_\ast^2}{d^2}\,\int_0^\infty\,I_\nu\,\tilde{p}\,d\tilde{p},
\end{equation}

\noindent where $I_\nu$ is the emergent intensity as given by

\begin{equation}
I_\nu = S_0\,\left[1-e^{-\tau_C(\tilde{p})}\right].
\end{equation}

\noindent When the wind is optically thick, such that the excess emission
from the wind greatly exceeds the attentuated stellar emisison through
the wind, the radiative transfer has a well-known solution when there is
no clumping (Panagia \& Felli 1975; Wright \& Barlow 1975).  
When constant clumping is present ($m=0$),
the spectral energy distribution is unchanged, and the flux is simply
enhanced above that of a smooth wind (Nugis et al.\ 1998).

Ignace (2009) also showed that an analytic solution can result with
a power-law distribution in the clumping.  The following integral
relation will be found of general use in subsequent steps:

\begin{equation}
\int_0^\infty\,\left(1-e^{-ax^{\beta}}\right)\,x\,dx = \frac{1}{\beta}\,
	\Gamma\left(\frac{2}{\beta}\right)\,a^{2/\beta},
\end{equation}

\noindent where $\Gamma$ is the Gamma-function.

For the case at hand, the continuum optical depth is

\begin{eqnarray}
\tau_C(\tilde{p}) & = & \int_{-\infty}^{+\infty}\,{\cal T}_C\,
	\frac{d\tilde{z}}{\tilde{r}^{4+m}} \\
 & = & {\cal T}_C\,\tilde{p}^{-3-m}\,\int_0^\pi\,(\sin\theta)^{2+m}
	\, d\theta \\
 & = & {\cal T}_C\,\tilde{p}^{-3-m}\,G_{\rm m}(\pi),
\end{eqnarray}

\noindent where the second line above uses a change of variable
to $\theta$, with $\tan \theta = p/z$, and

\begin{equation}
G_{\rm m}(\theta) = \int_0^\theta\,(\sin x)^{2+m}\, dx.
\end{equation}

\noindent The flux of continuum emission becomes

\begin{equation}
f_C = \frac{2\pi R_\ast^2}{d^2}\,S_0\,\left(\frac{1}{3+m}\right)\,
	\Gamma\left(\frac{2}{3+m}\right)\,\left[
	G_{\rm m}(\pi)\,{\cal T}_C(\lambda)\right]^{2/(3+m)}.
	\label{eq:fC}
\end{equation}

\noindent In the Rayleigh-Jeans limit, the continuum flux
will have a power-law slope of $-2+4.2/(3+m)$, with $S_0$
scaling as $\lambda^{-2}$ for the Planck function.  When $m=0$,
the canonical slope of $-0.6$ results.  Formally, the analytic
solution of Equation~(\ref{eq:fC}) requires that $m > -1$.

Within the line, the solution is really not any more complicated.
Again, Ignace (2009) showed that

\begin{equation}
f(w_{\rm z}) = \frac{2\pi R_\ast^2}{d^2}\,S_0\,\int_0^\infty\,
	\left\{1 - \exp\left[-({\cal T}_C\,G_{\rm m}(\pi)+{\cal T}_L\sin\theta)
	\tilde{p}^{-3-m}\right]\right\}\,\tilde{p}\,d\tilde{p}.
\end{equation}

\noindent While the argument of the exponential now has two terms,
the form of the integral is just like that of the 
pure continuum.  The analytic result is

\begin{equation}
f(w_{\rm z}) = \frac{2\pi R_\ast^2}{d^2}\,S_0\,\left(
	\frac{1}{3+m}\right)\,\Gamma\left(\frac{2}{3+m}\right)\,
        \left[G_{\rm m}(\pi)\,{\cal T}_C(\lambda)
	+ {\cal T}_L\,\sin\theta\right]^{2/(3+m)}.
\end{equation}

\noindent Note that $\sin\theta = \sqrt{1-w_{\rm z}^2}$.  When $m=0$,
the result of Rodr\'{i}guez (1982) is recovered.  The foremost outcome
for when the line and continuum source functions are equal is
that regardless of the value of $m$, the line profile is always
symmetric about line center.  However, when the two source functions
are not equal, the line shape will be asymmetric, as demonstrated
in the next section.

\subsection{Solution for the Case of $S_L \ne S_C$}
\label{sub:different}

In the previous section, a relatively complicated radiative transfer
problem for line and continuum was found to have an analytic solution.
The simplifications required to obtain that solution were spherical
symmetry, time-independent flow, and the limit of constant wind expansion.
Variation in the clumping factor could be included if the variation
can be treated as a power law.  Especially key was that both the
free-free and line opacities scaled as the square of density.

The final key assumption was that the line and continuum source
functions were equal.  However, this assumption can be relaxed
to allow for unequal source functions (yet still constant throughout
the flow at large radius).  In this case the solution for the 
emergent intensity is more complicated, and becomes

\begin{equation}
I_\nu(\tilde{p},w_{\rm z}) = S_C\,\left(e^{-\tau_W}-e^{-\tau_C}\right)\,
	e^{-\tau_L} + S_C \, \left(1-e^{-\tau_W}\right)
	+ S_L\,\left(1-e^{-\tau_L}\right)\,e^{-\tau_W}.
\end{equation}

\noindent This expression has three terms.  A ray at impact
parameter $\tilde{p}$ intersects the conical isovelocity zone
in the form of a ring.  Considering just one point on this ring
corresponding to $\tilde{z}$ for a given velocity shift $w_{\rm z}$,
we have two path segments and one point to consider for the accumulation
of sinks and sources that contribute to the emergent intensity.
The first term in the expression is for the continuum emission
up to the point of interest, and then its attenuation by the line
opacity at the point.  The second term is for the continuum emission
from the point of interest to the observer.  The third term is
the contribution by the line emission, as attenuated by the foreground
continuum opacity.  Thus as before, $\tau_C$ is the total continuum optical
depth along the ray, and we also have $\tau_W$ as the continuum
optical depth from the observer to the point of interest where the line
emissivity contributes to the emission.  

When $S_L=S_C$, terms involving $\tau_W$ cancel out.  With 
unequal source functions, the dependence on $\tau_W$ persists.
The emergent intensity now becomes:

\begin{equation}
I_\nu(\tilde{p}) = S_C\,\left(1-e^{-\tau_C-\tau_L}\right)
	- (S_L-S_C)\,e^{-\tau_W-\tau_L} + (S_L-S_C)\,e^{-\tau_W}.
	\label{eq:Inu}
\end{equation}

\noindent For this expression the first term closely mimics the
result from the preceding section when $S_L=S_C$.  Thus the other
two terms in the arrangement of Equation~(\ref{eq:Inu})
represent modifications when the source functions
are unequal.

The flux still has an analytic solution.  However, an additional
standard integral relation is required, of the form

\begin{equation}
\int_0^\infty x^{-\beta}\,e^{-ax}\,dx = \Gamma(1-\beta)\,a^{\beta-1}.
\end{equation}

\noindent This relation can be applied to the solution for the flux
by allowing $x=\tilde{p}^{-3-m}$, for which $\tilde{p} = x^{-1/(3+m)}$.
One also has $pdp = -(3+m)\,x^{-\beta}\,dx$ with $\beta= (5+m)/(3+m)$.

The flux in the continuum, outside the velocities of the line, is the
same as in the preceding section.  However, within the line, the
flux now becomes

\begin{eqnarray}
\frac{f(w_{\rm z})}{f_0} & = &
	\left(\frac{1}{3+m}\right)\,\Gamma\left(\frac{2}{3+m}\right)\,
        \left[G_{\rm m}(\pi)\,{\cal T}_C(\lambda)
	+ {\cal T}_L\,(\sin\theta)^{1+m}\right]^{2/(3+m)} \nonumber\\ 
 & & 	- \delta_{LC}\,\Gamma\left(\frac{-2}{3+m}\right)\,
	\left[G_{\rm m}(\theta)\,{\cal T}_C(\lambda)
        + {\cal T}_L\,(\sin\theta)^{1+m}\right]^{2/(3+m)} \nonumber\\ 
 & & 	+ \delta_{LC}\,\Gamma\left(\frac{-2}{3+m}\right)\,
	\left[G_{\rm m}(\theta)\,{\cal T}_C(\lambda)\right]^{2/(3+m)},
\end{eqnarray}

\noindent where

\begin{equation}
\delta_{LC} = \frac{S_L}{S_C} - 1,
\end{equation}

\noindent and

\begin{equation}
f_0 = \frac{2\pi R_\ast^2}{d^2}\,S_C.
\end{equation}

It is frequently the case that line profile data are plotted as
continuum normalized.  The continuum-normalized emission line profile
is given by

\begin{eqnarray}
\frac{f(w_{\rm z})}{f_C} & = &
        \left[1+\frac{t_{LC}}{G_{\rm m}(\pi)}(\sin\theta)^{1+m}
	\right] \nonumber\\ 
 & &	+ \delta_{LC}\,\gamma_{\rm m}\,\left\{
	\left[\frac{G_{\rm m}(\theta)}{G_{\rm m}(\pi)}
        + \frac{t_{LC}}{G_{\rm m}(\pi)}\,(\sin\theta)^{1+m}
	\right]^{2/(3+m)} \right. \nonumber\\ 
 & &	\left. -\left[\frac{G_{\rm m}(\theta)}{G_{\rm m}(\pi)}
	\right]^{2/(3+m)}\right\}
	\label{eq:norm}
\end{eqnarray}

\noindent where

\begin{equation}
t_{LC}(\lambda) = {\cal T}_L / {\cal T}_C,
\end{equation}

\noindent and

\begin{equation}
\gamma_{\rm m} = -(3+m)\,\frac{\Gamma[-2/(3+m)]}{\Gamma[+2/(3+m)]},
\end{equation}

\noindent where the negative anticipates that the Gamma function
in the numerator
is also negative.

Note the following special cases.  Of course, where there
is no line opacity, the radio SED
will be a power law in wavelength with 

\begin{equation}
f_\nu \propto B_\nu\,{\cal T}_c^{2/(3+m)} \propto g_\nu^{2/(3+m)}\,
	\lambda^{-2(1+m)/(3+m)}.
\end{equation}

\noindent for $S_C = B_\nu(T_C)$.
The opposite extreme is when the line opacity is significant,
but the continuum is negligible.  The emission line profile shape
becomes

\begin{equation}
f(w_{\rm z}) \propto S_L\,({\cal T}_L)^{2/(3+m)}\,\left(1
	-w_{\rm z}^2\right)^{1/(3+m)},
\end{equation}

\noindent Note that the line shape is symmetric about line center
in the limit of a strong line.
Wavelength dependence pertinent to the specific line transition is
implied through the factors $S_L$ and ${\cal T}_L$.

\begin{figure*}[!t]
\includegraphics[width=\columnwidth]{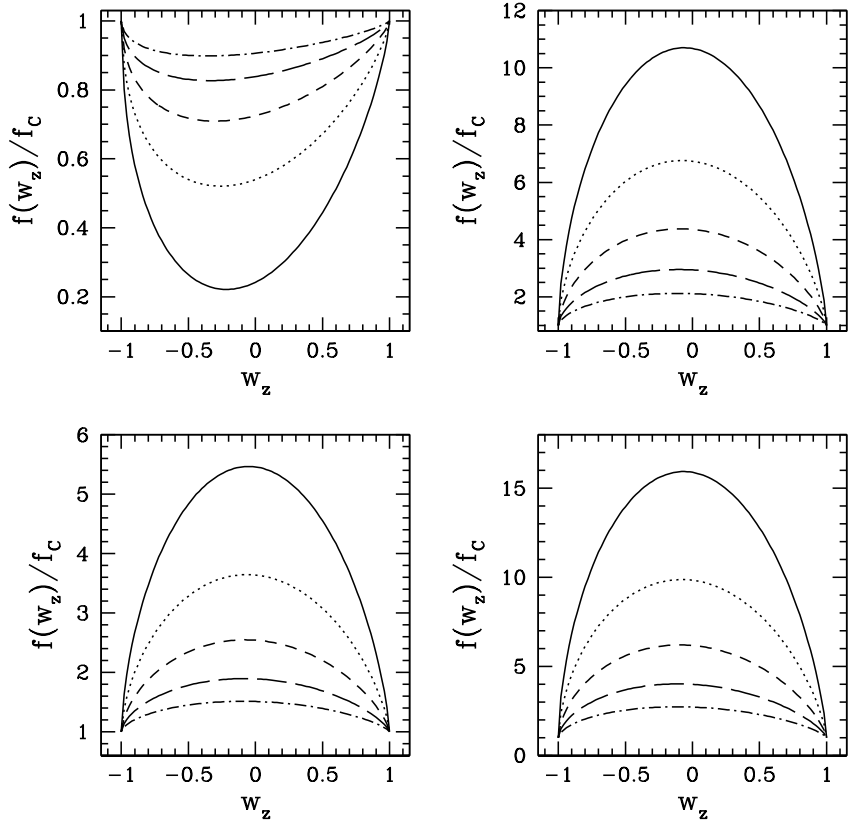}
\caption{Continuum normalized emission line profiles for the case of
$m=-0.5$.  The 4 panels are:  upper left is for $\delta_{LC} = -0.15$,
lower left is for $\delta_{LC} = +0.23$, upper right is for $\delta_{LC}
= +0.62$, and lower right is for $\delta_{LC} = +1.0$.  In each panel, the
5 line profiles are for $t_{LC} = 0.32$ (dot-dashed), 0.75 (long dashed),
1.8 (short dashed), 4.2 (dotted), and 10 (solid), from the weakest line
to the strongest.  Lines are plotted against normalized velocity shift.
Note that each panel has a different scale.}
\label{fig2}
\end{figure*}

\begin{figure*}[!t]
\includegraphics[width=\columnwidth]{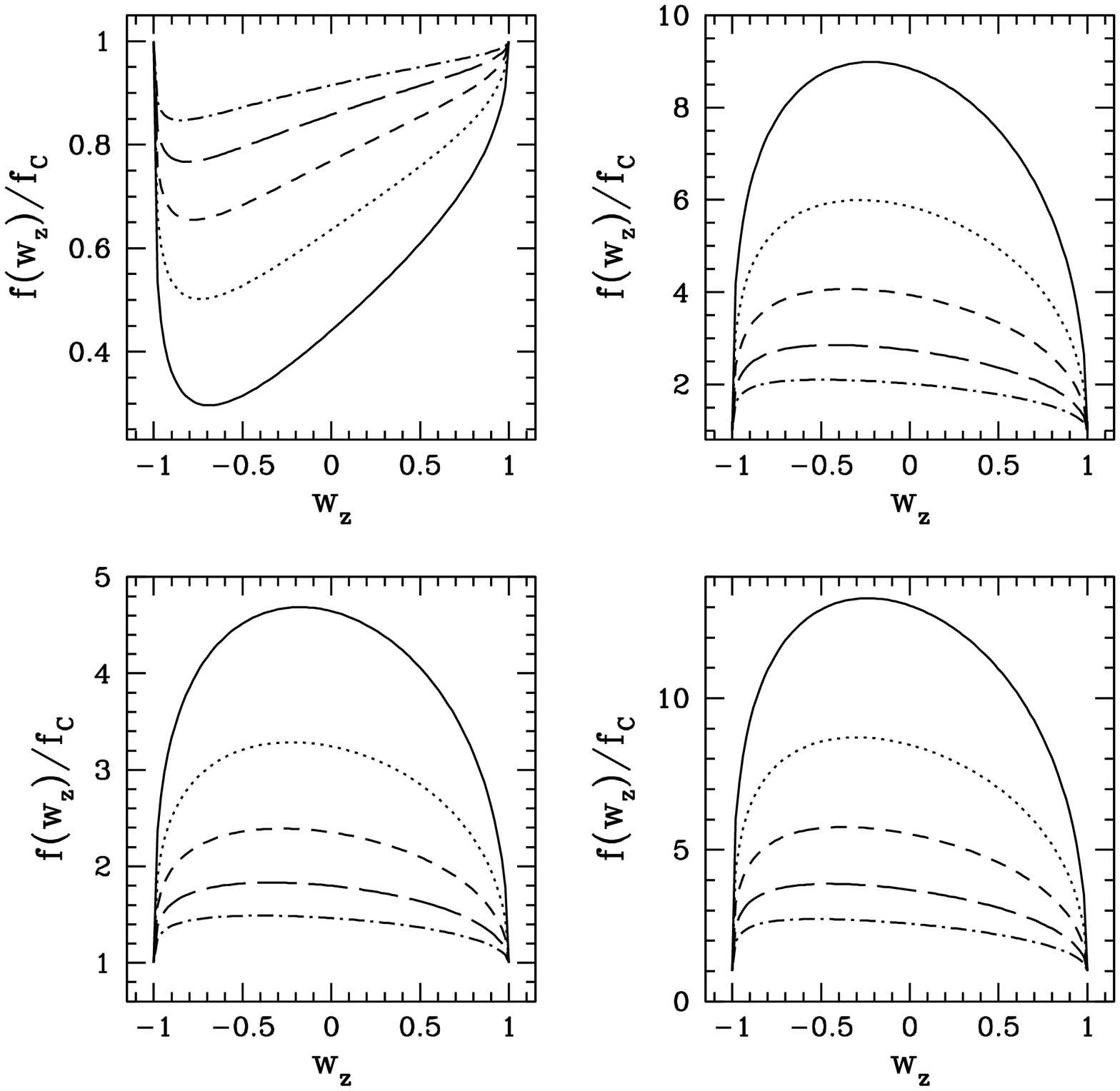}
\caption{As with Fig.~\ref{fig2}, except for $m=0$.}
\label{fig3}
\end{figure*}

Equation~(\ref{eq:norm}) is the main result of this study.  
The first term represents a symmetric component to the emission
line profile.  The subsequent two terms contribute generally
to asymmetric influences to the line in the form
of $G_{\rm m}(\theta)$.  These influences depend on the clumping
power-law exponent $m$, on the ratio of the source functions
$S_L/S_C$, and on the ratio of optical depths ${\cal T}_L / {\cal
T}_C$.  Note that if $\delta_{LC} > 0$, the line is in emission,
whereas for $\delta_{LC} < 0$, the line is in absorption.  Illustrative
examples are given in the following section.  

\section{Results}
\label{sec:results}

\begin{figure*}[!t]
\includegraphics[width=\columnwidth]{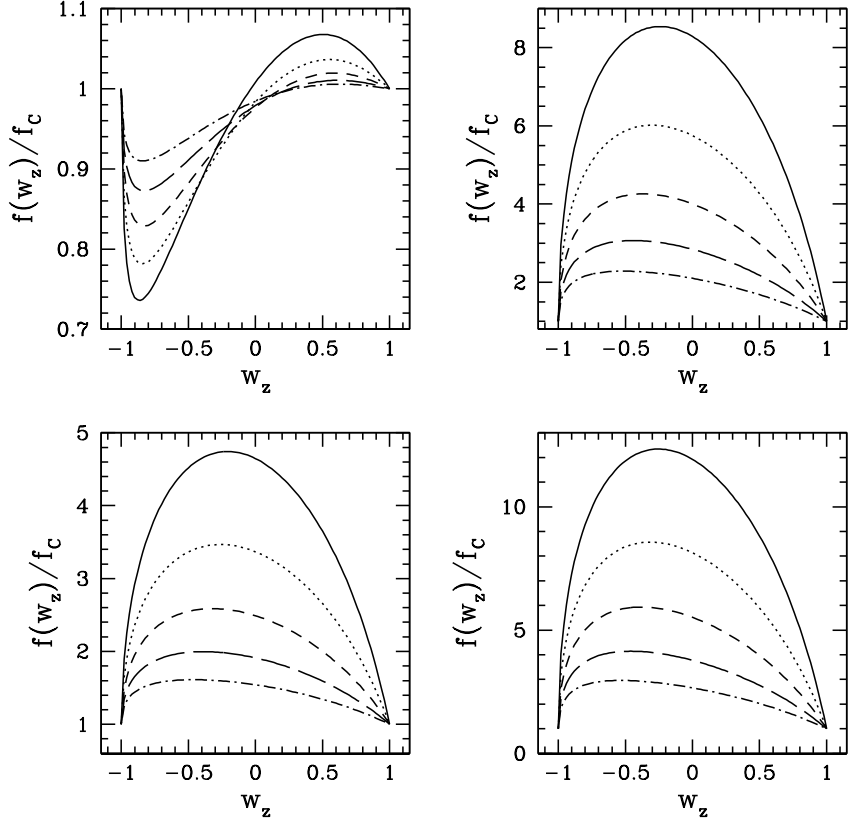}
\caption{As with Fig.~\ref{fig2}, except for $m=1$
and with $\delta_{LC} = -0.1, 0.4, 0.9$, and 1.4.}
\label{fig4}
\end{figure*}

Figures~\ref{fig2}--\ref{fig4} provide illustrative results for line
profile shapes.  Figure~\ref{fig2} is for $m=-0.5$ (i.e., clumping that
increases with radius); Figure~\ref{fig3} is for $m=0$ (i.e., clumping
that is constant with radius); and Figure~\ref{fig4} is for $m=+1$
(i.e., clumping that declines with radius).  Each figure has 4 panels:
upper left is for $\delta_{LC} = -0.15$, lower left is for $\delta_{LC}
= +0.23$, upper right is for $\delta_{LC} = +0.62$, and lower right
is for $\delta_{LC} = +1.0$ for Figures~\ref{fig2} and \ref{fig3},
but $\delta_{LC} = -0.1, 0.4, 0.9,$ and 1.4 for Figure~\ref{fig4}.
Each panel has 5 line profiles, with $t_{LC} = 0.32, 0.75, 1.8, 4.2,$
and 10, from the weakest line to the strongest.  The profiles are
continuum normalized and plotted against velocity shift, $w_{\rm z}$.
Note that each panel has a different ordinate scale.

In the case that $\delta_{LC}<0$, the line profile is actually 
in absorption. In the case of $m=+1$, the line shape takes the
appearance of a weak P~Cygni line shape, with blueshifted net
absorption and redshifted net emission.  

For $m=-0.5$, the line profiles are more symmetric about line
center as compared to either $m=0.0$ or $m=+1.0$.  As $m$ approaches $-1$, the line profile becomes perfectly
symmetric, because the factor contributing to the line asymmetry
cancels exactly when $2/(3+m) = 1$.  As $m$ increases,
the line shapes become increasingly asymmetric with the line skewed
preferentially toward blueshifted velocities.  This is natural generally
because the attentuation of line emission from the farside hemisphere of
the wind is greater than from the nearside hemisphere.  When $S_L=S_C$,
absorption is exactly compensated by emission, and no asymmetry in
the line can result.  For the given assumptions,
the line asymmetry occurs only when the source
functions are unequal.

\begin{figure}[!t]
\includegraphics[width=\columnwidth]{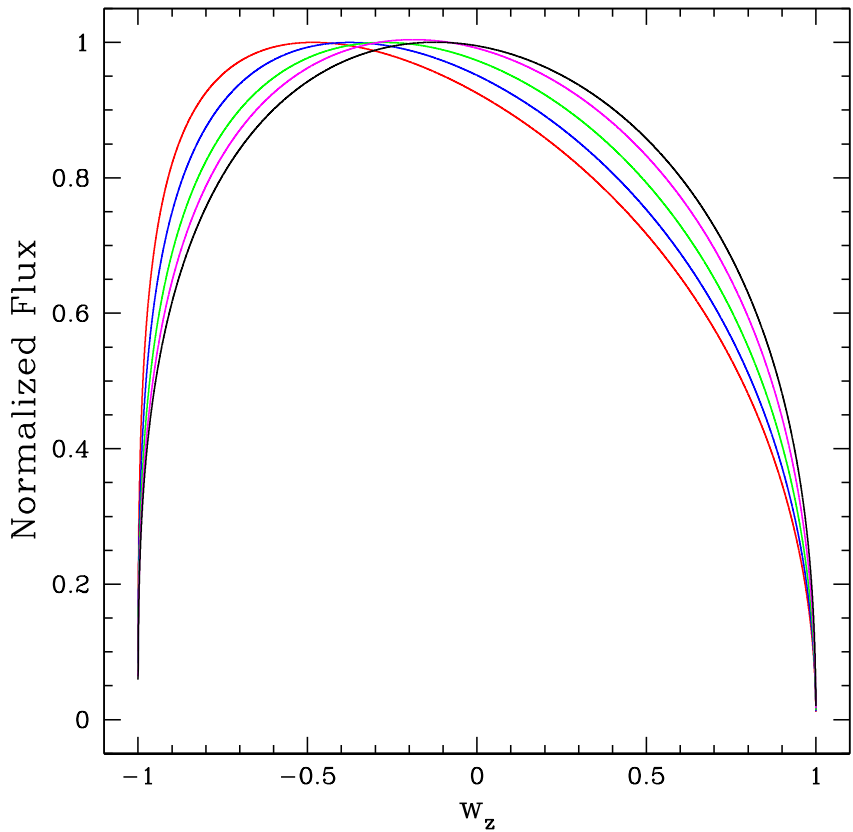}
\caption{The line profiles for the case $m=0.0$ shown
to highlight the evolution of line asymmetry as a function
of $t_{LC}$.  The line profiles have been continuum subtracted and
normalized to peak emission of unity.  The parameter $\delta_{LC}=0.5$
was held fixed.  The line profiles are for
$\log t_{LC} = -0.5$ (red), $-0.125$ (blue),
$+0.25$ (green), $+0.625$ (magenta), and $+1.0$ (black) 
}
\label{fig5}
\end{figure}

The line profiles display some degeneracy between $\delta_{LC}$
and $t_{LC}$.  As $t_{LC}$ becomes large, asymmetry in the line shape
lessens, in the sense that the peak emission shifts closer to line
center.  A large line optical depth means that (positive) $\delta_{LC}$
has less influence on the line shape.  Generally, $t_{LC}$ controls
the degree of asymmetry in the line, and $\delta_{LC}$ acts as an
overall amplitude for the line emission (or line equivalent width).

\section{Conclusions}
\label{sec:concs}

The focus of this contribution has been to highlight the asymmetry
of RRLs arising from a spherical wind
using an analytic derivation.  Previous analytic work produced
symmetric line shapes.  Numerical calculations have demonstrated
that asymmetric lines can be produced.  Here, with the assumption
of constant but unequal line and continuum source functions,
asymmetric line shapes are produced.  The derivation allows for the
presence of microclumping in the wind in terms of a power-law
distribution (rising or declining with radius from the star).  While
the clumping distribution can impact line asymmetry, the line asymmetry
results even with constant clumping, or no clumping whatsoever.
Under the model assumptions, emission line asymmetry arises from the
continuum opacity (specifically the appearance of the term with
$G_{\rm m}(\theta)$ in Equation~[\ref{eq:norm}]) that absorbs the
redshifted emission from the far hemisphere more than the blueshifted
emission from the near hemisphere.  The result is a line shape with
blueshifted emission peak.  (The same effect arises in X-ray lines;
c.f., Ignace 2016b).

RRLs are vigorously pursued as a diagnostic of source properties,
from kinematics to geometrical aspects.  Peters, Longmore, \&
Dullemond (2012) have made an indepth study of various factors
that affect the flux of line emission and the shape of the line profile,
including line asymmetry.  Observational motivation for understanding
line asymmetry of RRLs include some objects as the early-type binary
MWC349, specifically the H76$\alpha$ line (Escalante et al.\ 1989).
Understanding the line formation is key for distinguishing between
radiative transfer effects for the line formation versus the influence
of aspherical effects intrinsic to the source, such as binarity or
asymmetric mass-loss or flow geometry.  Applications for such effects
include emission-line objects like MWC349 (e.g., LkH$\alpha$ 101;
Thum et al.\ 2013), outflow from star-forming clumps (e.g., Kim et
al.\ 2018), and Planetary Nebulae (e.g., Ershov \& Berulis 1989;
S\'{a}nchez Contreras et al.\ 2017).  Analytic solutions are valuable
to these studies in two main respects:  (a) they allow for rapid
evaluation of parameter space that can be honed with more detailed
numerical calculations to fit data, and (b) they are important
for providing non-trivial benchmarks against which numerical codes
can be tested.

\acknowledgements

The author expresses appreciation to an anonymous referee for
several helpful comments.

\end{document}